\documentclass[twocolumn,showpacs,amssymb,prd]{revtex4}

\usepackage{graphicx}
\usepackage{dcolumn}
\usepackage{bm}
\usepackage{epsfig}

\begin{document}
\newcommand{\gsim}{\hbox{\rlap{$^>$}$_\sim$}}
\newcommand{\lsim}{\hbox{\rlap{$^<$}$_\sim$}}


\title{A Common Solution of Two Cosmic Puzzles}

\author{Shlomo Dado} \affiliation{Department of Physics and Space Research
Institute, Technion, Haifa 32000, Israel}

\author{Arnon Dar}
\affiliation{Department of Physics and Space Research
Institute, Technion, Haifa 32000, Israel}

\begin{abstract} 
The origin of the high energy gamma-ray background, which was 
measured with the large area telescope (LAT) aboard the Fermi satellite at 
energy below 820 GeV, and of the diffuse cosmic background of 
neutrinos, which was observed at much higher energies with the IceCube 
detector deep under the south pole ice, are among the current unsolved 
major cosmic puzzles. Here we show that their properties indicate a common 
origin: the decay of mesons produced in collisions of cosmic rays 
accelerated in relativistic jets with matter inside these jets. Moreover, 
their properties are those expected if the the highly 
relativistic jets are those emitted mainly in core 
collapse supernovae of type Ic, and by Active Galactic nuclei.
\end{abstract}

\maketitle 

\section{Introduction} 

The high energy gamma ray background radiation (GBR) was first discovered 
36 years ago with the SAS-2 satellite$^1$ and was later studied in detail 
with the Compton Gamma Ray Observatory$^2$ (CGRO) and more recently with 
the Fermi gamma ray satellite$^3$. Recently, also a high energy neutrino 
background radiation (NBR) was discovered with the megaton IceCube 
detector$^4$ deep under the ice of the south pole. 
High energy particle physics offers three common production mechanisms of 
both the GBR and NBR. They include production of mesons in hadronic 
collisions of cosmic rays (CRs) with ambient matter in the galactic 
interstellar medium$^5$ (ISM) and in the intergalactic medium$^6$ (IGM) 
whose decays produce photons, electrons and neutrinos, photo production of 
mesons in CR collisions with radiation in/near gamma ray source$^7$, and 
decay of massive dark matter particle relics from the Big Bang$^8$. But, 
so far, no connection has been found$^9$ between the NBR and GBR. 
and the origin and observed properties  are still unsolved cosmic puzzles. 
Here we 
show that the very high energy NBR discovered$^1$ with IceCube is that 
expected from the GBR observed with Fermi-LAT at much lower energies$^3$, 
if both were produced by hadronic collisions of cosmic rays accelerated in 
relativistic jets with matter inside these jets, which are launched mostly 
in supernovae explosions of type Ic that produce the long duration gamma 
ray bursts$^{10,11}$ (GRBs), and by active galactic nuclei (AGN).

\section{Hadronic production of high energy gamma rays and neutrinos} 
The highly relativistic jets launched in core collapse SN 
explosions of type Ic that produce long duration GRBs 
and by AGN also accelerate 
through the Fermi mechanism$^{12}$ the swept in ionized matter in front of 
them$^{10,11}$ to cosmic ray energies with a power-law 
spectrum$^{12}$ $dn/dE\propto E^{-{2}}$. These cosmic rays escape the jets 
by diffusion in the turbulent magnetic fields inside the jets.  For a 
Kolmogorov spectrum$^{13}$ of the random magnetic fields in the jets, the 
CRs' diffusion coefficient is proportional to $E^{1/3}$ and consequently the 
residence time of the CRs in the jets is $t_r(E)\propto E^{-1/3}$.
Hadronic collisions of these high energy CRs in the jets  
produce mesons (mainly $\pi$'s and $K$'s), which decay shortly 
after to the stable particles: photons ($\gamma$), electrons ($e^\pm$) and 
neutrinos ($\nu_i, \bar{\nu}_i$ of the three known flavors $i=e,\mu,\tau$). 
The decay $\pi^0\rightarrow 2\gamma$ with energy per photon 
$E_\gamma=m_{\pi^0}/2\approx$ 67.5 MeV in the $\pi^0$ rest-frame dominates 
$\gamma$-ray production. Neutrino production is dominated by the cascade 
decay of $\pi^\pm\rightarrow\mu^\pm\nu_\mu\rightarrow \ e^\pm 
\nu'_e\nu_\mu\bar{\nu}_\mu$, where 
$\nu'_e$ stands for electron neutrino in $\pi^+$ decay and electron 
antineutrino in $\pi^-$ decay, and where the $e^\pm$ and the three neutrinos 
have each an energy $<\!E\!>\approx E_\pi/4\approx$ 35 MeV in the $\pi^\pm$ 
rest frame.

When a high energy CR proton of energy $E_p$ produces a $\pi$ with a lab 
frame energy $E_\pi=x\, 
E_p$, Lorentz invariance implies that the average lab frame energy of the 
2 $\gamma$'s from $\pi^0$ decay is $(x/2)\,E_p $ whereas that of the 
$\nu's$ is $\approx (x/4)\, E_p$. High energy collisions satisfies 
approximately Feynman scaling$^{14}$, namely, the $x$-distribution of the 
produced mesons does not depend on $E_p\,.$  Since the CR residence time 
in the jets is proportional to $E^{-1/3}$, the spectra of 
the produced high energy mesons, photons and neutrinos become $\propto 
E^{-2.33}$.  Inelastic pp collisions produce roughly equal numbers of 
$\pi^0$'s, 
$\pi^+$'s and $\pi^-$'s with the same $x$-distribution, and neutrino 
oscillations convert the produced neutrino flux into three equal fluxes of 
neutrinos of the three neutrino flavors. Thus, the produced $\gamma$-ray 
flux and neutrino flux per flavor due to CR production of $\pi$ mesons 
satisfy the simple relation
\begin{equation}  
\Phi_\nu \approx  \left[{m_{\pi^\pm} \over 
2\,m_{\pi^0}}\right]^{1.33}\Phi_\gamma=0.42\, \Phi_\gamma.
\end{equation}
Eq.~(1) is valid 
for the injected spectrum of $\gamma$-rays and 
neutrinos when meson production in hadronic collisions
of the high energy  cosmic rays takes place inside the
highly relativistic jets 
launched by the main cosmic accelerators (supernovae, pulsars, compact 
binaries, microquasars, and blazars).
However, the observed spectrum of the GBR is
modified by absorption of extragalactic high energy $\gamma$-rays 
in  $\gamma+\gamma_{EBL}\!\rightarrow\!e^+e^-$ 
collisions with the photons of the extragalactic background light (EBL). 
Moreover, the EGB has been observed below 820 GeV, while
the NBR has been  detected  in the 35 Tev - 3 PeV energy 
range (where the background of atmospheric neutrinos becomes 
negligible). Consequently, Eq.~(1) cannot be tested directly 
because the NBR and EGB were measured in different energy ranges.
 
\section{Are the NBR and GBR related?}
Above 100 GeV the flux of the extra galactic photons is progressively 
absorbed by the EBL$^{15}$. This attenuation of the EGB, however, becomes 
small at$^{16}$ $E\lsim 100$ GeV and the unattenuated EGB can be 
extrapolated from there to the 35 TeV- 3 PeV energy range where it can be 
compared to the NBR that has been measured$^4$ with IceCube. The lack of a 
precise knowledge of the EBL$^{17}$ does not allow a precise theoretical 
evaluation of the attenuation of the EGB. But, presumably, the observed 
attenuated EGB can be parametrized as a power-law that represents the 
unattenuated EGB with an exponential cutoff, which represents the 
attenuation. Indeed, the EGB that was measured during 50 months$^3$ with 
the Fermi Large Area Telescope (Fermi-LAT) is very well fit 
($\chi^2/ndof=7.05/23$) by 
\begin{equation} E^2\Phi_\gamma=6.4\times 
10^{-7} \left[{E\over {\rm GeV}}\right]^{-0.30}\, {\bf e^{-{E\over 
0.366{\rm TeV}}}}\,{{\rm GeV\over cm^2\, sr\,s}}, 
\end{equation} 
as shown in Fig.~1.  
\begin{figure}[] \centering \vspace{-2cm} 
\epsfig{file=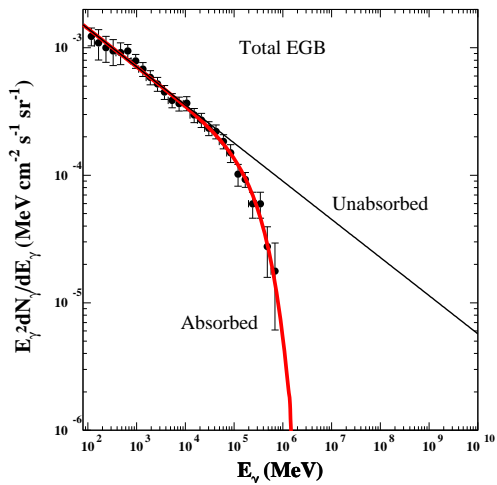,width=7.cm,height=7.cm} 
\caption{The sub TeV flux of 
the extragalactic gamma-ray background (EGB) as function of gamma-ray 
energy measured$^3$ with Fermi-LAT, and the best fit exponential cutoff 
power-law behavior as given in Eq.~2. The straight line represents the 
unattenuated power-law EGB.} 
\label{Fig1} 
\end{figure} 
The attenuation of the EGB decreases its contribution to the GBR flux at 
$E_0=100$ GeV by $\approx 0.45\times 10^{-11}\,{\rm GeV/cm^2\, sr\,s}$. 
Adding this loss to the mean flux of the GBR observed with Fermi-LAT$^4$ 
at 100 GeV yields unattenuated sky averaged GBR flux, 
\begin{equation} 
\Phi_\gamma(GBR)\approx 
3.95 \times 10^{-11}\, \,{\rm GeV \over cm^2\, sr\,s}\,. 
\end{equation}

The contribution of bremsstrahlung and inverse Compton 
scattering of background photons by cosmic ray electrons and 
cosmic ray production of of $\pi^0$ in the ISM 
decreases rapidly with energy. Hence, if the unattenuated 
GBR at E$>100$ GeV is dominated by hadronic meson production, then
the $\nu/\gamma$ production ratio yields a $\nu$ flux per flavor
\begin{equation}
E^2\,\Phi_\nu\approx 0.42\, E^2 \Phi_\gamma
=2.1\times 10^{-8}
           \left[{E\over 100{\rm TeV}}\right]^{-0.30} 
           {{\rm GeV\over cm^2\, sr\, s }}.   
\end{equation}
The power-law index is harder than 1/3 probably because of the 
increase with energy of the total inelastic pp cross section. 
About $\sim41\%$ of the flux is  extragalactic in origin 
and $\sim 59\%$ Galactic in origin. 
Moreover, 
the sky distribution of the GBR and NBR are expected to be nearly
identical. 
Both, the magnitude and the spectral index of this  predicted
flux agrees  with  the best fitted single power-law, 
$E^2 \Phi_\nu \approx  1.4\times 10^{-8}\,
           [E/ 100\,{\rm TeV}]^{-0.30}\,
           {\rm GeV\, cm^{-2}\, s^{-1}\, sr^{-1}}$
reported$^4$ recently by the IceCube collaboration.
However, since  $<E_\nu>\approx <E_\pi/4>\approx 0.06\,E_p$
and the flux of CR nucleons has a knee around ${\rm E=2}$ PeV,
where its power-law index changes$^{}$ from $\approx$ -2.7 to $-3$,
the spectral index of the NBR is predicted to change around 150 TeV from 
-2.3 to -2.6, 
\begin{equation}
E^2\,\Phi_\nu\approx  
0.42\, E^2 \Phi_\gamma
=1.85\times 10^{-8}
           \left[{E\over 150{\rm TeV}}\right]^{-0.60}
           {{\rm GeV\over cm^2\, sr\, s }},
\end{equation}
for E$>150$ TeV, as shown in Fig.~2.
\begin{figure}[]
\centering 
\vspace{-2cm} 
\epsfig{file=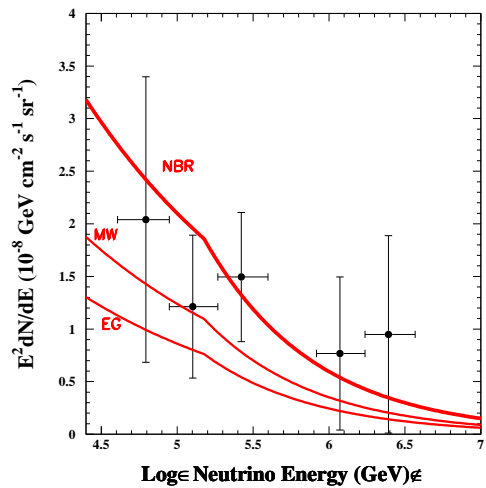,width=7.cm,height=7.cm} 
\caption{ 
Comparison between the flux of the high energy  neutrino 
background radiation (NBR) above 35 TeV measured$^4$ with IceCube and 
the flux expected from the measured GBR and EGB with Fermi-LAT (Fig.~1), 
assuming hadronic production of mesons mostly inside the CR sources.} 
\label{Fig2}
\end{figure}
The striking consistency between the 
GBR and NBR (in magnitude and spectral index) shown in Fig 2. is obtained 
neither for meson photo production in/near 
source nor for decay of massive relic particles from the 
Big-Bang (see Supplementary Material). 

\section{Which cosmic accelerators produce the EGB and ENB?} 
Among the cosmic accelerators that have been identified through their 
non-thermal radio, X-ray and $\gamma$-ray emissions, only the highly 
relativistic jets that produce the long duration gamma ray bursts in 
supernova explosions of type Ic and the jets of active galactic 
nuclei (AGN) appear to be able to accelerate cosmic ray 
nuclei to their highest observed energies with their observed spectra, 
composition and fluxes. The jets have a mean   
column density sufficiently large for efficient hadronic production 
of mesons by the CRs accelerated inside the jets.  

Upon launch, the highly relativistic jets emitted by SN type Ic explosions 
and by AGN are optically thick$^{18}$, mainly due to Compton scattering. 
Their expansion makes them transparent to radiation when their column 
density decreases below $ N_s\sim 1/\sigma_T\approx 1.5\times 10^{24}\, 
cm^{-2}$. The swept in ionized particles in front of them 
are Fermi accelerated by their turbulent magnetic fields  
and escape by diffusion in through these fields$^{18}$.
During their diffusion 
in the jets,, the CRs  suffer inelastic collisions with a probability 
$P\sim \sigma_{in}\,n_s\,c\,\tau_{esc} \propto E^{-\delta}$,
where $\delta\approx 0.30$ for E$<150$ TeV and
$\delta\approx 0.60$ for E$> 150$ TeV. They  
produce a broken power-law flux of secondary neutrinos and gamma-rays, with 
a spectrum  ${\rm \propto E^{-2+\delta}}$,  
with a probability 
that exceeds  by far the probability of being produced in  
the ISM of the host galaxy.

An estimate of the GRB 
contribution to the ENB, 
from such a source that is based on the very 
successful cannonball model of GRBs is$^{18}$,
\begin{equation}
E^2\Phi_\nu \approx {\dot{N}_{GRB}\over N_{GRB}}{m_p\over m_e}
\Sigma_i {f_\nu(z_i)\,(1\!+\!z_i)^{2/3} Eiso(z_i)
\over 12\, \pi\, [D_L(z_i)]^2}\,\left[{E\over
m_p}\right]^{-1/3}\,.
\end{equation}
In this estimate,  $\dot{N}_{GRB}$ is the observed rate of  
GRBs, $N_{GRBs}$ is the number of GRBs todate
with known redshift $z_i$, luminosity distance $D_L(z_i)$ and isotropic 
equivalent 
gamma ray energy $Eiso(i)$ and the summation extends over such GRBs
where $f_\nu(z_i)$ is the probability
that a CR proton that was accelerated in a GRB at 
redshift $z_i$, produced a neutrino in  
hadronic collisions of the  CRs in the jet. 
Using only priors and completely avoiding  adjustable  parameters, 
Eq.~(6) yields a neutrino flux$^{19}$  compatible (within errors)      
with our estimated  Galactic and extragalactic  contributions to the NBR.

\section{Conclusions} 
The diffuse gamma ray background that has been measured below 
820 GeV with Fermi-LAT and the diffuse neutrino background that was 
discovered with IceCube in the 35 TeV - 3 PeV energy range are consistent 
with 
being produced jointly in hadronic collisions of extragalactic cosmic rays 
in/near source. Such a common source is very likely the highly relativistic 
launched in supernovae type Ic, and by AGN, 
most of which do not point in our direction.
The emission from the jets may be spread over 
long time window and therefore may not be in 
conflict with the IceCube limits on the neutrino fluxes from 
individual$^{20}$ and stacked$^{21}$ GRBs, since they were derived from 
observations within rather a short time window covering the prompt 
emission and the early afterglow phases of GRBs (such a  short time window
is expected in the popular fireball models of GRBs$^{22}$).

The observed gamma ray and neutrino backgrounds are 
compatible with those predicted by the cannonball model of GRBs$^{22}$.  

Detailed calculations of cosmic ray production of gamma rays and neutrinos 
in hadronic collisions of galactic cosmic rays in the interstellar medium of 
our Galaxy and  external galaxies,  yield only a small 
contribution 
to the observed GBR and NBR with a much steeper spectral decline$^{19}$.

Although the approximate isotropy, spectral index and flux level of the
high energy neutrino background measured with IceCube ({\it 4}) are
consistent with an extragalactic origin, a dominant contribution to the
observed background from cosmic ray production of neutrinos in the Milky
Way with nearly the same sky distribution as the GBR
cannot be ruled out at present by the IceCube observations, because of
low statistics and the large uncertainty in the neutrino arrival
directions of most of the IceCube events.

\section{Supplementary Material}

\subsection{Photo mesons cannot produce the observed 
extragalactic $\gamma$-ray bacground}

The typical energy of photons $\epsilon'_\gamma$ in the rest frame 
(indicated by a prime) of sources, such as the highly relativistic 
jets of blazars or GRBs, is  less than a few keV. Consequently,
the typical threshold energy of CR protons for $\pi$ production in 
head-on collisions with photons in the rest frame of such sources  is
\begin{equation}
E'_{th}\approx {m_p(m_p+2m_\pi)\over 4\epsilon'_\gamma}\approx 283\, 
{{\rm keV}\over \epsilon'_\gamma}\, {\rm TeV}.
\end{equation}
Hence, if the typical bulk motion Lorentz factor of the  source is
$\Gamma$, then the typical energy 
of the photons from $\pi^0$ decay in CR sources is
\begin{equation}
E_\gamma\geq {1\over 2} {m_{\pi^0}\, \Gamma\, E'_{th}\over m_p+m_{\pi^0}}
\approx 8.8\,\Gamma\,\left[{{\rm keV}\over \epsilon'_\gamma}\right]\,
{\rm TeV}\, .
\end{equation}
We conclude that photoproduction of mesons by CRs cannot be a common 
source of the sub TeV EGB observed with Fermi-Lat below 820 GeV 
and the astrophysical neutrinos discovered with IceCube at $E>100 TeV.$

\subsection{Upper bounds on the ENB}
If the observed flux of cosmic rays at energies above the CR 
ankle at $E\sim 3$ EeV is 
extragalactic in origin, then it must be dominated by protons because at 
such 
energies extragalactic cosmic ray  nuclei are fully dissociated in 
collisions with the extragalactic background photons$^1$. 
Assuming that such CR protons produce at most a single $\pi^+$ 
meson at source, Waxman and Bahcall$^2$  derived  "a model-independent 
upper bound" ${E_\nu}^2 \Phi_\nu < 2\times 10^{-8}\, {\rm GeV/cm^2\, s\, sr}$
for the energy flux of extragalactic neutrinos above 100 TeV. 

The locally observed flux of ultrahigh energy cosmic rays (UHECRs)
indeed has a break at 40 EeV, which coincides in 
energy with that predicted by Greisen, Zatsepin and Kuzmin$^3$ (GZK) 
for extragalactic CR protons. 
The measured flux between the CR ankle and the GZK cutoff 
is well represented by$^4$ $\Phi_p\approx 0.16\,(E/GeV)^{-2.7}\, 
{\rm GeV^{-1}\, cm^{-1}\, s^{-1}\, sr^{-1}}$. (This spectrum also  
represents well, within errors,  the measured spectrum of CR protons and 
bound nucleons  below 
the CR ankle, at least down to to $\sim$ 100 PeV$^5$.
Such an extragalactic flux of CR nucleons can produce an 
extragalactic flux of high energy neutrinos through the decay
of $\pi^+$ mesons produced by collisions of the cosmic ray nucleons
with photons in/near their sources. If $\pi^+$  production  
is dominated by $p\gamma\rightarrow \Delta^+(1232)\rightarrow n\pi^+$,   
the produced $\pi^+$  carries on average a fraction 
$E'_\pi/M_\Delta\approx 0.22 $ 
of the incident proton momentum, where $E'_\pi$ is the $\pi$ energy 
in the $\Delta^+$ rest frame (this fraction for  $\pi^+$ production  near 
threshold  is $m_\pi/(m_\pi+m_n)\approx 0.13$). 
Since the branching ratio of
$\Delta(1232)\!\rightarrow\! n\pi^+$ is $\approx\! 1/2$, and $\pi^+$ decay
produces 3 neutrinos with  $<\!E_\nu\!>\approx m_\pi/4$,
production of neutrinos through
photoproduction of $\Delta^+(1232)$ by high energy cosmic ray 
protons in optically thin sources satisfies
\begin{equation}
\Phi_\nu \approx {3\over 2}\left[{0.216\over 4}\right]^{1.7}\, P 
\,\Phi_p,
\end{equation}
where $P=\sigma_{p\gamma}\, N_\gamma$  is the 
probability that a CR proton produces a $\pi^+$ 
in a proton collision with a photon column density $N_\gamma$ of 
an optically thin source before entering the intergalactic space. 
Eq.~(7) with a constant $P$, which is independent of source and redshift, 
does not 
depend explicitly on cosmic evolution. Hence, for an optically thin 
sources with $P=1$,  
the maximal energy flux per neutrino flavor is
\begin{equation}
E^2\Phi_\nu \approx 5.5\times 10^{-7}\, P\, \left[{E\over 100{\rm 
TeV}}\right]^{-0.7} {\rm {GeV\over cm^2\, s \, sr}}.
\end{equation}
For  P=1, this bound  is larger by 1.7 orders of magnitude than the
bound derived by Waxman and Bahcall who have   
assumed a power-law index -2 for 
the spectrum of the extragalactic flux of cosmic ray protons, while the 
observed index above and below the CR ankle is -2.70. Such an index  
increases by {\it two orders of magnitude!}
the estimated energy flux of the extragalactic CRs in the relevant energy 
range  $10^{16}-10^{18}$ eV  from the observed energy flux in the 
$10^{19}-10^{21}$ eV energy range.  Moreover,   
a constant $P$ yields an extagalactic 
neutrino flux proportional to the extragalactic flux of CR protons.  
This automatically includes the correct 
redshift dependence and cosmic evolution 
in the estimated neutrino flux.

\end{document}